\documentclass{ws-ijmpa}
\usepackage[super,compress]{cite}
\usepackage{graphicx}
\begin{document}
\markboth{Z.E. Musielak}{New Equation of Nonrelativistic Physics and 
Theory of Dark Matter}

\catchline{}{}{}{}{}
%
%%%%%%%%%%%%%%%%%%%%%%%%%%%%%%%%%%%%%%%%%%%%%%%%%%%%%%%%%%%%%%%%%%%%

\title{New Equation of Nonrelativistic Physics and 
Theory of Dark Matter}

\author{Z.E. Musielak}

\address{Department of Physics, The University of Texas at 
Arlington, Arlington, TX 76019, USA\\
zmusielak@uta.edu}

\maketitle

\begin{history}
\received{Day Month Year}
\revised{Day Month Year}
\end{history}

\begin{abstract}
Two infinite sets of Galilean invariant equations are derived using 
the irreducible representations of the orthochrous extended Galilean 
group.  It is shown that one set contains the Schr\"odinger equation, 
which is the fundamental equation for ordinary matter, and the other 
set has a new asymmetric equation, which is proposed to be the 
fundamental equation for dark matter.  Using this new equation, 
a theory of dark matter is developed and its profound physical 
implications are discussed.  This theory explains the currently known 
properties of dark matter and also predicts a detectable gravitational 
radiation.
\keywords{Galilean groups; Galilean invariance; equations of physics;
dark matter}
\end{abstract}

%\ccode{PACS numbers:}

\section{Introduction}

There is strong astronomical evidence for the existence of Dark Matter 
(DM) in the Universe [1-3].  DM is a concept whose physical meaning
is currently unknown despite numerous theories that have attemped to
explain it.  The theories have postulated different particles [4-6] but so 
far none of these particles have been found experimentally [7-9].  To
overcome this impass, a novel theoretical approach is presented.  In 
this approach, a new equation of non-relativistic physic is discovered 
and this paper reports on a theory of DM based on this equation.

Searching for new equations is conducted within the framework of 
Galilean space and time.  The Galilean Principle of Relativity (GPR) 
is used to define a class of Galilean observers [10].  According to 
GPR, all Galilean observers agree that physical laws are the same 
in all inertial frames of reference.  Moreover, the observers also 
obey the Principle of Analyticity (PA) [11], which requires that 
only analytic ($\mathcal{C}^{\infty}$) functions are used.

Dynamical equations describing the space and time evolution of a
scalar wavefunction are derived using the irreducible representations 
(irreps) of the orthochronous extended Galilean group [12].  The 
derived equations can be classified as symmetric and asymmetric 
(in space and time) equations with constant coefficients.  It is shown
that none of the symmetric equations is Galilean invariant and that 
among all asymmetric equations, there are only two infinite sets 
of Galilean invariant equations.  One of these sets contains {\it 
Schr\"odinger-like equations}, and the elements of the other set 
are newly discovered second-order equations, which are called 
here the {\it new asymmetric equations}.  

The prominent feature of these two sets of the Galilean invariant 
equations is their dependence on constant coefficients, which 
must be determined.   In this paper, an equation is called {\it 
fundamental} if, and only if, it is local, Galilean invariant, its 
Lagrangian exists and its coefficient is determined uniquely 
by the properties of either Ordinary Matter (OM) or Dark 
Matter (DM).  

Among the two sets of derived equations, one fundamental 
equation for each set was identified.  It is shown that one 
of the Schr\"odinger-like equations becomes the fundamental 
Schr\"odinger equation for OM, and that one of the new 
asymmetric equations is proposed to be considered as the 
fundamental equation for DM.  If this choice of the fudamental 
equation for DM is correct, it has significant physical implications 
as it requires a new constant of Nature valid for DM only.  The 
constant is called the {\it quanta of energy}, and its presence 
suggests that quantization of the energy of DM is frequency 
independent.  The physical meaning of this new constant is 
discussed and a model of DM is constructed.  It is also 
suggested that the fundamental equation for DM, with 
interactions included, may be used to formulate a quantum 
theory of DM, which would become the non-relativistic limit 
of a future relativistic theory of DM.

The paper is organized as follows: in Section 2, Galilean groups 
and the eigenvalue equations are described;  symmetric and 
asymmetric equations are given in Section 3; Galilean invariant 
equations are presented in Section 4;  the fundamental equations 
for ordinary and dark matter are identified in Section 5; and 
conclusions are given in Section 6.

\section{Galilean groups and eigenvalue equations}
 
The Galilean group of the metric $\mathcal {G}$ is the group of 
Galilean space and time, and its structure is $\mathcal {G} = [T(1) 
\otimes O(3)] \otimes_s [T(3) \otimes B(3)]$, where $T(1)$, $O(3)$, 
$T(3)$ and $B(3)$ are subgroups of translations in time, rotations, 
translations in space, and boosts, respectively [12].  The general 
structure of the extended Galilean group is $\mathcal {G}_e = 
[O(3) \otimes_s B(3)] \otimes_s [T(3+1) \otimes U(1)]$, where 
$T(3+1)$ is an invariant Abelian subgroup of combined translations 
in space and time, and $U(1)$ is a one-parameter unitary subgroup 
[13].  Note that $\mathcal {G}_e$ does not include the space and 
time inversions.  Observers who use the group $\mathcal {G}_e$ 
are called the extended Galilean observers and they must also 
agree on the square of the wavefunction in all intertial frames [12]. 

The group $\mathcal {G}_e$ is known to have scalar and spinor 
irreducible representations (irreps), which are physical; that is, 
they allow uniquely defining elementary particles by scalar 
(Schr\"odinger equation) [10] or spinor (L\'evy-Leblond equation) 
[13,14] wavefunctions.  The vector and tensor representations are
not physical because they do not allow particles to be localized [12].  
The invariant subgroup $T(3+1) = T(3) \otimes T(1) \subset 
\mathcal{G}_e$ has the irreps, and a scalar wavefunction $\phi$ 
transforms as one of these irreps if, and only if, the following 
eigenvalue equations [10,15] are satisfied 
\begin{equation}
i {{\partial} \over {\partial t}} \phi (t, \mathbf {x}) = \omega \phi 
(t, \mathbf {x})\ ,
\label{eq1}
\end{equation}  
and 
\begin{equation}
- i \nabla \phi (t, \mathbf {x}) = \mathbf {k} \phi (t, \mathbf {x})\ ,
\label{eq2}
\end{equation}  
where $\omega$ and $\mathbf {k}$ are labels of $\phi (t, \mathbf {x})$,
which is an eigenfunction of the generators of the invariant Abelian 
subgroup $T(3+1)$.

\section{Local symmetric and asymmetric equations}

In general, the equations derived from the eigenvalue equations can be 
divided into two separate families, namely, the symmetric equations, with
the same order of space and time derivatives, and the asymmetric equations,
with different orders of space and time derivatives.  Since only local (either 
first or second-order) equations are derived, the higher derivative equations 
[15] are not considered in this paper.

By combing the eigenvalue equations (Eqs. \ref{eq1} and \ref{eq2}), the 
following symmetric frst-order equation is obtained
\begin{equation}
\left [ {{\partial} \over {\partial t}} -  C_{s} {\mathbf k} \cdot \nabla \right ] 
\phi (t, \mathbf {x}) = 0\ , 
\label{eq3}
\end{equation}  
where $C_s = \omega / k^{2}$.  Since the labels $\omega$ and ${\mathbf k}$
of the irreps may have any real value, there is an infinite number of $C_s$ and 
first-order equations.   

Similarly, the symmetric second-order equation can be written as  
\begin{equation}
\left [ {{\partial^2} \over {\partial t^2}} -  C_{w} \nabla^{2} \right ] 
\phi (t, \mathbf {x}) = 0\ , 
\label{eq4}
\end{equation}  
where $C_w = \omega^2 / k^{2}$ with $k^{2n} = (\mathbf {k} \cdot \mathbf 
{k})^{n}$.  With $C_w$ being arbitrary, there is an infinite set of second-order 
equations.  In a special case $C_w = v^2_w$, where $v_w$ = const is a
characteristic wave speed, Eq. (\ref{eq4}) becomes the regular wave equation
of classical physics.  

There are also two infinite sets of asymmetric second-order equations:   
\begin{equation}
\left [ i {{\partial} \over {\partial t}} + C_{s} \nabla^{2} \right ] 
\phi (t, \mathbf {x}) = 0\ ,
\label{eq5}
\end{equation}  
and 
\begin{equation}
\left [ {{\partial^2} \over {\partial t^2}} -  i  C_{w} {\mathbf k} \cdot 
\nabla \right ] \phi (t, \mathbf {x}) = 0\ .
\label{eq6}
\end{equation}  
Following [10,15], who originally derived the infinite set of equations 
given by Eq. (\ref{eq5}), these equations are called the Schr\"odinger-like 
equations.  However, Eq. (\ref{eq6}) represents a set of {\it new asymmetric 
equations} derived in this paper.  

All local equations that can be obtained from the eigenvalue equations, 
which are based on the irreps of the orthochrous extended Galilean group, 
are given by Eqs. (\ref{eq3}) through (\ref{eq6}).  Since the derivation 
does not guarantee that these equations are Galilean invariant, their 
invariance must be verified.

\section{Galilean invariant equations}

\subsection{Symmetric equations}

Having derived the local symmetric and asymmetric equations, their 
invariance with respect to the Galilean transformations of $\mathcal 
{G}_e$, which involve translations in space and time, rotations and 
boosts, is now verified.  

Let $S$ and $S^{\prime}$ be two intertial frames of reference in 
Galilean space and time, and let $\mathbf {v}$  = const be the 
velocity with which $S^{\prime}$ moves with respect to $S$.  
Then, a boost is given by $\mathbf {x} = \mathbf {x}^{\prime} 
+ \mathbf {v} t$ and $t^{\ \prime}\ =\ t$, which can be written as 
\begin{equation}
{{\partial} \over {\partial t}} = {{\partial} \over {\partial t^{\prime}}} 
- \mathbf {v} \cdot \nabla^{\prime}\ ,
\label{eq7}
\end{equation}  
and $\nabla = \nabla ^{\prime}$ [16].  Applying these transformations 
to the eigenvalue equations (Eqs. \ref{eq1} and \ref{eq2}), it is seen 
that the transformed equations have different forms, which means that 
they are not Galilean invariant.  

It can be shown that all symmetric and asymmetric equations are invariant 
with respect to the Galilean translations and rotations, which are the subgroups 
of $\mathcal {G}_e$.  However, their boost invariance must be investigated 
independently by using the following transformation law for the wavefunction 
$\phi (t, \mathbf {x}) = \chi (t^{\prime}, {\mathbf {x}}^{\prime}) 
\phi^{\prime} (t^{\prime}, {\mathbf {x}}^{\prime})$, where $\chi (t^{\prime}, 
{\mathbf {x}}^{\prime})$ is to be determined [15].  The results previously 
obtained show that none of the symmetric equations given by Eqs.  (\ref{eq3}) 
and (\ref{eq4}) is Galilean invariant [15].

\subsection{Schr\"odinger-like equations}

Galilean invariance of the Schr\"odinger-like equations given by Eq. (\ref{eq5}) 
requires that the coefficient $C_{s}$ is the same as the coefficient $C^{\prime}_{s}$ 
in the following transformed Schr\"odinger-like equations  
\begin{equation}
\left [ i {{\partial} \over {\partial t^{\prime}}} + C^{\prime}_{s} 
\nabla^{{\prime} 2} \right ] \phi^{\prime} (t^{\prime}, \mathbf 
{x}^{\prime}) = 0\ ,
\label{eq8}
\end{equation}  
where $C^{\prime}_{s} = \omega^{\prime} / k^{{\prime} 2}$.
 
The addional requirement for the Galilean invariance is that the 
wavefunction $\phi (t, \mathbf {x})$ transforms as   
\[
\phi (t, \mathbf {x})\ =\ \phi (t^{\prime}, {\mathbf {x}^{\prime}} + 
\mathbf {v} t^{\prime})\ =\ \phi^{\prime} ({\mathbf {x}^{\prime}}, 
t^{\prime})\ \chi (t^{\prime}, {\mathbf {x}}^{\prime}) 
\]
\begin{equation}
\hskip0.5in = \phi^{\prime} ({\mathbf {x}^{\prime}}, t^{\prime})\ 
e^{i \eta (t^{\prime}, {\mathbf {x}^{\prime}})} ,  
\label{eq9}
\end{equation}
where the phase factor $\eta (t^{\prime}, {\mathbf {x}^{\prime}}) = 
\left ( \mathbf {v} \cdot \mathbf {x}^{\prime} + v^2 t^{\prime} / 2 
\right ) / (2 C^{\prime}_{s})$ [16,17]; this standard phase factor 
for Galilean transformations appears also in the Galilean 
invariant Newton laws of dynamics [18] from which it can be 
removed in a self-consistent way [19]. 

The above two conditions are necessary for all Schr\"odinger-like 
equations with arbitrary $C_{s}$ to be Galilean invariant, which 
means that all extended Galilean observers agree upon the form 
of the dynamical equation and its wavefunction in their inertial 
frames of reference.

In previous work [10,15], it was assumed that $C_{s} = 
C^{\prime}_{s}$ in all inertial frames; a formal proof of this 
is now presented in Proposition 1.  

{\bf Proposition 1:}  The Schr\"odinger-like equations given by 
Eq. (\ref{eq5}) are Galilean invariant if, and only if, $C_{s} = 
\omega / k^{2} = C^{\prime}_{s} = \omega^{\prime} / 
k^{{\prime} 2}$ in all inertial frames of reference.   

\noindent
{\bf Proof:} The general solutions to Eq. (\ref{eq5}) 
can be written as 
\begin{equation}
\phi (t, \mathbf {x})\ = \phi_o e^{- i ( \omega t - \mathbf {k} \cdot 
\mathbf {x})}\ ,
\label{eq10}
\end{equation}
and
\begin{equation}
\phi^{\prime} (t^{\prime}, \mathbf {x}^{\prime})\ = \phi^{\prime}_o 
e^{- i ( \omega^{\prime} t^{\prime} - \mathbf {k}^{\prime} \cdot 
\mathbf {x}^{\prime})}\ ,
\label{eq11}
\end{equation}
where $\phi_o$ and $\phi^{\prime}_o$ are constant amplitudes.  
Combining these solutions with Eq. (\ref{eq9}) and assuming that 
the amplitudes are the same, $\phi_o = \phi^{\prime}_o$, then, 
the result yields 
\begin{equation}
\mathbf {k} \cdot \mathbf {x} - \omega t = \mathbf {k}^{\prime} 
\cdot \mathbf {x}^{\prime} - \omega^{\prime} t^{\prime} + 
{1 \over {2 C^{\prime}_s}} \mathbf {v} \cdot \mathbf {x}^{\prime}
+ {1 \over {4 C^{\prime}_s}} v^2 t^{\prime}\ .
\label{eq12}
\end{equation}

Using the Galilean transformations $t^{\prime} = t$ and $\mathbf 
{x}^{\prime} = \mathbf {x} - \mathbf {v} t$, and also substituting 
$C^{\prime}_s = C_s$, Eq. (\ref{eq12}) becomes
\begin{equation}
\left ( \mathbf {k}^{\prime} - \mathbf {k} + {1 \over {2 C_s}} 
\mathbf {v} \right ) \cdot \mathbf {x} + \left ( \omega  - \omega^{\prime} 
- \mathbf {k}^{\prime} \cdot \mathbf {v} + {1 \over {4 C_s}} 
v^2 \right ) t = 0\ ,
\label{eq13}
\end{equation}
which gives 
\begin{equation}
\mathbf {k}^{\prime} = \mathbf {k} - {1 \over {2 C^{\prime}_s}} 
\mathbf {v}\ , 
\label{eq14}
\end{equation}
and, taking this result into account, the expression for $\omega^{\prime}$
becomes 
\begin{equation}
\omega^{\prime} = \omega  - \mathbf {k} \cdot \mathbf {v}
+ {1 \over {4 C^{\prime}_s}} v^2\ . 
\label{eq15}
\end{equation}

Therefore, 
\begin{equation}
{\omega^{\prime} \over \mathbf {k}^{{\prime} 2}} =
{ {\omega  - \mathbf {k} \cdot \mathbf {v} + v_o^2 / {4 C_s}}
\over {k^2 - \mathbf {k} \cdot \mathbf {v} / C_s + ( v / 2 C_s )^2}} 
= {\omega \over k^2}\ , 
\label{eq16}
\end{equation}
and 
\begin{equation}
\left ( {{\omega} \over {C_s}} - k^2 \right ) \left ( {{v^2} \over 
{4 C_s}} - \mathbf {k} \cdot \mathbf {v} \right ) = 0\ .
\label{eq17}
\end{equation}
Since the second term on the LHS of this equation is non-zero,
then $C_s = \omega / k^2$.  This conludes the proof.

{\bf Corollary 1:}  The value of $C_s$ in the Schr\"odinger-like equations
(Eq. \ref{eq5}) has the same value in all inertial frames of reference. 

According to Corollary 1, the constant $C_s$ plays the same role in
Galilean Relativity as the speed of light in the Special Theory of Relativity;  
this means that $C_s$ remains the same for all Galilean observers.   

The main result is that there is an infinite set of Schr\"odinger-like equations
that are Galilean invariant and that they only differ by their different values 
of $C_s$.  The significance of this constant in the physics of OM and DM is 
discussed in Sect. 5.

\subsection{New asymmetric equations} 

The new asymmetric equations given by Eq. (\ref{eq6}) also form 
an infinite set because the constant $C_w$ may have any real value. 
Typically, Galilean invariance requires a phase factor in the wavefunction 
[13,15-17].  However, the results of the following proposition show 
that no phase factor cannot be defined to make these asymmetric 
equations Galilean invariant.

{\bf Proposition 2:}  Let $\phi (t, \mathbf {x})$ be the wavefunction 
of Eq. (\ref{eq8}), $\phi^{\prime} ({\mathbf {x}^{\prime}}, 
t^{\prime})$ be the transformed wavefunction, and $\eta (t^{\prime}, 
{\mathbf {x}^{\prime}})$ its phase factor given by Eq. (\ref{eq9}).
Then, no $\eta (t^{\prime}, {\mathbf {x}^{\prime}})$ exists to 
guarantee Galilean invariance of the new asymmetric equations.

\noindent
{\bf Proof:} After performing the Galilean transformations, Eq. (\ref{eq6}) 
becomes
\[
\left [ {{\partial^2} \over {\partial t^{\prime 2}}} -  i  C^{\prime}_{w} 
{\mathbf k^{\prime}} \cdot \nabla^{\prime} \right ] \phi
(t^{\prime}, \mathbf {x^{\prime}}) 
\]
\begin{equation}
\hskip0.25in - \left [ 2 ( {\mathbf v} \cdot \nabla^{\prime} ) {{\partial} 
\over {\partial t^{\prime}}} - ( {\mathbf v} \cdot \nabla^{\prime} )^2 
\right ] \phi (t^{\prime}, \mathbf {x^{\prime}}) = 0\ ,
\label{eq18}
\end{equation}  
This equation has two extra terms when compared to Eq. (\ref{eq6}), 
and therefore it is not Galilean invariant.  

Introducing the phase factor $\eta (t^{\prime}, {\mathbf {x}^{\prime}})$ 
in Eq. (\ref{eq9}) and some algebric manipulations, Eq. (\ref{eq18}) can 
be written as 
\[
\left [ {{\partial^2} \over {\partial t^{\prime 2}}} -  i  C^{\prime}_{w} 
{\mathbf k^{\prime}} \cdot \nabla^{\prime} \right ] \phi^{\prime} 
(t^{\prime}, \mathbf {x^{\prime}}) e^{i \eta (t^{\prime}, {\mathbf 
{x}^{\prime}})}
\]
\begin{equation}
\hskip0.25in - \left [ 2 ( {\mathbf v} \cdot \nabla^{\prime} ) {{\partial} 
\over {\partial t^{\prime}}} - ( {\mathbf v} \cdot \nabla^{\prime} )^2 
\right ] \phi^{\prime} (t^{\prime}, \mathbf {x^{\prime}}) e^{i \eta 
(t^{\prime}, {\mathbf {x}^{\prime}})} = 0\ ,
\label{eq19}
\end{equation}  
which is of the same form as Eq. (\ref{eq18}), except the exponential 
term with the phase factor.  The two extra terms remain; thus, there is 
no $\eta (t^{\prime}, {\mathbf {x}^{\prime}})$ that can make Eq. 
(\ref{eq19}) Galilean invariant or of the same form as Eq. (\ref{eq6}).
This conludes the proof.

{\bf Corollary 2:}  The results of Proposition 2 can be generalized to 
$\phi (t, \mathbf {x})\ =\ \phi^{\prime} ({\mathbf {x}^{\prime}}, 
t^{\prime})\ \chi (t^{\prime}, {\mathbf {x}^{\prime}})$, where
$\chi$ is any differentiable function. 

Since Galilean invariance cannot be guaranteed by introducing a 
phase factor, a different method is now presented in the following
proposition.

{\bf Proposition 3:}  The new asymmetric equation given by Eq. 
(\ref{eq6}) becomes Galilean invariant if, and only if, $\phi 
({\mathbf {x}}, t) = \phi ({\mathbf {r}})$ and $\phi^{\prime} 
({\mathbf {x}^{\prime}}, t^{\prime}) = \phi^{\prime} 
({\mathbf {r}^{\prime}})$, where ${\mathbf {r}} = \mathbf {x}
+ \mathbf {v} t / 2$ and ${\mathbf {r}^{\prime}} = {\mathbf 
{x}^{\prime}} + {\mathbf {v}} t^{\prime} / 2$. 

\noindent
{\bf Proof:} Comparison of Eq. (\ref{eq6}) and Eq. (\ref{eq19}) 
shows that these equations are of the same form if
\begin{equation}
\left [ 2 ( {\mathbf v} \cdot \nabla^{\prime} ) {{\partial} \over 
{\partial t^{\prime}}} - ( {\mathbf v} \cdot \nabla^{\prime} )^2 
\right ] \phi^{\prime} (t^{\prime}, {\mathbf x^{\prime}}) = 0\ ,
\label{eq20}
\end{equation}
where $\phi^{\prime} (t^{\prime}, {\mathbf {x}^{\prime}}) = 
\phi^{\prime} ({\mathbf {r}^{\prime}})$ and $\eta (t^{\prime}, 
{\mathbf {x}^{\prime}}) = 0$.  Using Eq. (\ref{eq20}), it can 
be shown that Eq. (\ref{eq19}) is of the same form as Eq. 
(\ref{eq6}).  

It must be also noted that after $\phi ({\mathbf {x}}, t) = \phi 
({\mathbf {r}})$ is substituted into Eq. (\ref{eq6}), its form 
remains the same as that of Eq. (\ref{eq19}); this concludes 
the proof.

The results of Proposition 3 can be used to derive the following 
equation for $\phi ({\mathbf {x}}, t) = \phi ({\mathbf {r}})$
\begin{equation}
{{d^2 \phi} \over {d ({\mathbf k} \cdot {\mathbf r})^2}} - i
C_{w,v} {{d \phi} \over {d ({\mathbf k} \cdot {\mathbf r})}}
= 0\ ,
\label{eq21a}
\end{equation}
or after the integration 
\begin{equation}
{{d \phi} \over {d ({\mathbf k} \cdot {\mathbf r})}} - i
C_{w,v} \phi  = C_0\ ,
\label{eq21b}
\end{equation}
where ${\mathbf {r}} = \mathbf {x} + \mathbf {v} t / 2$,
$C_{w,v} = 4 C_{w} / v^2$ and $C_0$ is an integration 
constant. 

Similarly, the corresponding equation for $\phi ({\mathbf {x}^{\prime}}, 
t^{\prime}) = \phi^{\prime} ({\mathbf {r}^{\prime}})$ becomes
\begin{equation}
{{d^2 \phi^{\prime}} \over {d ({\mathbf k^{\prime}} \cdot 
{\mathbf r^{\prime}})^2}} - i C^{\prime}_{w,v} {{d \phi^{\prime}} 
\over {d ({\mathbf k^{\prime}} \cdot {\mathbf r^{\prime}})}}
= 0\ ,
\label{eq22a}
\end{equation}
or after the integration
\begin{equation}
{{d \phi^{\prime}} \over {d ({\mathbf k^{\prime}} \cdot 
{\mathbf r^{\prime}})}} - i C^{\prime}_{w,v} \phi  = 
C^{\prime}_0\ ,
\label{eq22b}
\end{equation}
where ${\mathbf {r}^{\prime}} = {\mathbf {x}^{\prime}} + {\mathbf {v}} 
t^{\prime} / 2$, $C^{\prime}_{w,v} = 4 C^{\prime}_{w} / v^2$ and 
$C^{\prime}_0$ is an integration constant.  

Equations (\ref{eq21a}), (\ref{eq21b}), (\ref{eq22a}) and (\ref{eq22b}) 
are of the same form if, and only if, $C_{w,v}= C^{\prime}_{w,v}$ and 
$C_o = C^{\prime}_0$ have the same values for all extended Galilean 
observers.

Finding the solutions to Eqs. (\ref{eq21a}) and (\ref{eq22a}) is straigthforward;
they can be written as 
\begin{equation}
\phi ({\mathbf k} \cdot {\mathbf r}) = - i {{C_1} \over {C_{w,v}}} 
e^{i C_{w,v} ({\mathbf k} \cdot {\mathbf r})} + C_2\ ,
\label{eq23}
\end{equation}
and
\begin{equation}
\phi^{\prime} ({\mathbf k^{\prime}} \cdot {\mathbf r^{\prime}}) = 
- i {{C^{\prime}_1} \over {C^{\prime}_{w,v}}} e^{i C^{\prime}_{w,v} 
({\mathbf k^{\prime}} \cdot {\mathbf r^{\prime}})} + C^{\prime}_2\ ,
\label{eq24}
\end{equation}
where $C_1$, $C_2$, $C^{\prime}_1$ and $C^{\prime}_2$ are integration 
constants.  The solutions to Eqs. (\ref{eq21b}) and (\ref{eq22b}) are of the 
same form as those given by Eqs. (\ref{eq23}) and (\ref{eq24}) if, and 
only if, $C_2 = i C_0 / C_{w,v}$ and $C^{\prime}_2 = i C^{\prime}_0 / 
C^{\prime}_{w,v}$.   

For these solutions to be of the same form, it is required that 
$C_1 = C^{\prime}_1$, $C_2 = C^{\prime}_2$ and $C_{w,v} = 
C^{\prime}_{w,v}$.  The following proposition presents the necessary 
conditions to guarantee Galilean invariance of Eqs. (\ref{eq21a}) and 
(\ref{eq22a}).

{\bf Proposition 4:}  The conditions for the Galilean invariance $C_{w} = 
\omega^2 / k^{2} = C^{\prime}_{w} = \omega^{\prime 2} / 
k^{{\prime} 2}$, $C_1 = C^{\prime}_1$ and $C_2 = C^{\prime}_2$
are satisfied if, and only if  
\begin{equation}
{\mathbf k^{\prime}} = \xi (x,t) {\mathbf k}\ ,
\label{eq25}
\end{equation}
where
\begin{equation}
\xi (x,t) = \left ( {\mathbf x} + {1 \over 2} {\mathbf v} t \right )^2
\left ( x^2 - {1 \over 4} v^2 t^2 \right )^{-1}\ ,
\label{eq26}
\end{equation}
and $x = \vert {\mathbf x} \vert$.

{\bf Proof:} According to Proposition 3, there is no phase factor;
thus, $\phi ({\mathbf k} \cdot {\mathbf r}) = \phi^{\prime} 
({\mathbf k^{\prime} } \cdot {\mathbf r^{\prime} })$, and 
Eqs. (\ref{eq23}) and (\ref{eq24}) give
\begin{equation}
- i {{C_1} \over {C_{w,v}}} e^{i C_{w,v} ({\mathbf k} 
\cdot {\mathbf r})} + C_2 = - i {{C^{\prime}_1} \over 
{C^{\prime}_{w,v}}} e^{i C^{\prime}_{w,v} 
({\mathbf k^{\prime}} \cdot {\mathbf r^{\prime}})} 
+ C^{\prime}_2\ .
\label{eq27}
\end{equation}
Taking $C_1 = C^{\prime}_1$, $C_2 = C^{\prime}_2$
and $C_{w} = C^{\prime}_{w}$, Eq. (\ref{eq27}) simplifies 
to  
\begin{equation}
{\mathbf k} \cdot \left ({\mathbf x} + {1 \over 2} {\mathbf v} 
t \right ) = {\mathbf k^{\prime}} \cdot \left ({\mathbf x^{\prime}} 
+ {1 \over 2} {\mathbf v} t^{\prime} \right )\ .
\label{eq28}
\end{equation}
Using the Galilean transformations $\mathbf {x}^{\prime} = 
\mathbf {x} - \mathbf {v} t$ and $t^{\ \prime}\ =\ t$, 
Eq. (\ref{eq28}) becomes
\begin{equation}
{\mathbf k} \cdot \left ({\mathbf x} + {1 \over 2} {\mathbf v} 
t \right ) = {\mathbf k^{\prime}} \cdot \left ({\mathbf x} 
- {1 \over 2} {\mathbf v} t \right )\ ,
\label{eq29}
\end{equation}
or
\begin{equation}
{\mathbf k^{\prime}} =  \left ( {\mathbf x} + {1 \over 2} 
{\mathbf v} t \right )^2 \left ( x^2 - {1 \over 4} v^2 t^2 \right )^{-1}
{\mathbf k}\ .
\label{eq30}
\end{equation}
This concludes the proof.

{\bf Corollary 3:}  Since ${\mathbf v} = $ const, the condition 
$C_{w,v} = C^{\prime}_{w,v}$ reduces to $C_{w} = 
C^{\prime}_{w}$, or more specifically to 
\begin{equation}
C_w = {{\omega^2} \over {k^2}} = {{\omega^{\prime 2}}
\over {k^{\prime 2}}} = C^{\prime}_w\ .
\label{eq31}
\end{equation}

The obtained results demonstrate that the new asymmetric equations
are Galilean invariant if, and only if, the argument of the wavefunction
is $(\mathbf {x} + \mathbf {v} t / 2)$, and the wavevector and the
transformed wavevector are related to each other by Eq. (\ref{eq31}).
The method to make equations Galilean invariant presented in 
Propositions 4 and 5 can also be applied to other second-order 
equations.

\section{Physical implications}

\subsection{Fundamental equations of physics}

The main results of this paper are the infinite sets of Galilean 
invariant Schr\"odinger-like and new asymmetric equations
given by Eqs. (\ref{eq5}) and (\ref{eq6}), respectively, 
which are local and their Lagrangians are known.

The coefficients $C_s$ and $C_w$ in the Schr\"odinger-like 
and new asymmetric equations, respectively, are called here 
the Galilean constants as they are the same for all extended 
Galilean observers; in other words, they resemble the constant 
speed of light in Special Theory of Relativity but their physical 
meaning is different, as shown next.

Since the main differences between the equations in each set 
are their constant coeffcients, it is suggested that only after 
specific values of these constants are determined by physical 
properties of matter, such equations with fixed constants are 
called the {\it fundamental equations of physics}.  After 
introducing this new definition, the fundamental equations 
of physics for ordinary and dark matter are now identified.

\subsection{Fundamental equation for ordinary matter}

The two infinite sets of Galilean invariant equations for the scalar 
wavefunction cannot be used to describe classical particles of OM
as their motion is governed by Newton's equations, which are 
Galilean invariant, if the force in the second law of dynamics is 
also Galilean invariant [11,13,19].  Previous work (e.g., [20-22])
showed some applications of the Schr\"odinger equation to classical
waves.  Similar applications of the Schr\"odinger-like and new 
asymmetric equations are also possible but they will not be 
considered here.

Instead, this study concentrates on the microscopic structure of OM 
and the quantum description of elementary particles, which requires 
taking into account the wave-particle duality [17].  In the Galilean 
Relativity the energy-momentum relationship is given by $E = p^2 
/ 2m$, where $E$, $p$ and $m$ are the kinetic energy, momentum 
and mass of the free particles, respectively.  Based on this relationship, 
let $C_{s} = \omega / k^2 = \alpha_0 / 2m$, with $\alpha_0$ being a 
constant of Nature to be determined from the physical properties of OM. 

Applying the de Broglie relationship $\mathbf {p} = \hbar \mathbf {k}$ 
to $C_{s}$, one obtains 
\begin{equation}
\omega = {{\alpha_0 k^2} \over {2 m}} = {{\alpha_0 p^2} \over {2 m 
\hbar^2}} = {{\alpha_0} \over {\hbar^2}} E = {{\alpha_0} \over 
{\hbar}} \omega\ .
\label{eq32}
\end{equation}  
This expression is valid if, and only if, $\alpha_0 = \hbar$, or $\alpha_0$ 
is the same as the Planck constant.  Then, Eq. (\ref{eq7}) with $C_{s} 
= \alpha_0 / 2m = \hbar / 2m$ becomes the fundamental Schr\"odinger 
equation of Quantum Mechanics [17].

Since there are also infinitely many new asymmetric equations, it
remains to be determined whether one of them can also describe the
microscopic properties of OM; this problem is discussed in Sect. 5.3,
where the equations are applied to DM.

\subsection{Fundamental equation for dark matter}

As of today, no electromagnetic radiation of any kind has been detected 
from DM [4-9].  The NASA Bullet Cluster observations show a different 
behavior of DM when compared to OM [23].  This may imply that DM 
does not have the same quantum structure as OM and that the 
fundamental Schr\"odinger equation with $C_s = \hbar / 2 m$ is not 
the fundamental equation for DM; nevertheless, other Schr\"odinger-like 
equations with different $C_s$ are now explored.

Let $C_{s} = \alpha_1 / 2 m$, where $\alpha_1$ is a constant of Nature 
whose value is different than the Planck constant, namely, either $\alpha_1 
< \hbar$ or $\alpha_1 > \hbar$.  If this constant exists for DM, then the 
resulting Schr\"odinger-like equation would be the fundamental equation 
for DM.  The main problem with this idea is that the dimension of $\alpha_1$ 
is the same ($J \cdot s$) as the dimension $\alpha_0 = \hbar$ which, 
from a physical point of view, is {\it highly unlikely} that there are
two constants of Nature with identical dimensions but different values.  
Based on this argument, it is concluded that no other Schr\"odinger-like 
equation can become the fundamental equation for DM.

The only other possibility is that such fundamenatal equation for DM 
is among the set of new asymmetric equations discovered in this paper.  
The coefficient $C_{w,v}$ is given by
\begin{equation}
C_{w,v} = {{4} \over {v^2}} C_w = {{4} \over {v^2}} 
{{\omega^2} \over {k^{2}}}\ ,
\label{eq33}
\end{equation}
and for $C_w$ see Eq. (\ref{eq4}). 

For a DM non-relativistic elementary particle of mass $m$, the 
coefficient can be written as 
\begin{equation}
C_w = {{\omega^2} \over {k^2}} = {{\beta_0} \over {2m}}\ ,
\label{eq34}
\end{equation}
where $\beta_0$ is a constant of Nature for DM only.  Since 
$\beta_0$ is measured in $J$, which is different than the Planck
constant, let $\beta_0 = \varepsilon_o$ be a {\it quanta of 
energy of DM}.  Then 
\begin{equation}
C_{w,v} = {{\varepsilon_o} \over {2m}} {{4} \over {v^2}} =
{{\varepsilon_o} \over {2m}} {{2m} \over {\varepsilon_v}} = 
{{\varepsilon_o} \over {\varepsilon_v}}\ ,
\label{eq35}
\end{equation}
where $\varepsilon_v = m v^2 / 2$ is the kinetic energy of DM
particles confined to the intertial frame moving with the velocity 
$v = \vert {\mathbf v} \vert$.   The energy $\varepsilon_v$ 
normalizes the quanta of energy of DM, so that the coefficient 
in the fundamental equation for DM (see Eq. \ref{eq21a}) and 
in its solution (see Eq. \ref{eq23}) is dimensionless. 

Based on the above results, it is proposed that the {\it fundamental 
equation for DM is the new asymmetric equation} given by
\begin{equation}
\left [ {{\partial^2} \over {\partial t^2}} -  i  {{\varepsilon_o} 
\over {2m}} {\mathbf k} \cdot \nabla \right ] \phi (t, \mathbf {x}) 
= 0\ , 
\label{eq36}
\end{equation}  
whose Galilean invariant form is 
\begin{equation}
{{d^2 \phi} \over {d ({\mathbf k} \cdot {\mathbf r})^2}} - i
{{\varepsilon_o} \over {\varepsilon_v}} {{d \phi} \over 
{d ({\mathbf k} \cdot {\mathbf r})}} = 0\ ,
\label{eq37}
\end{equation}
and its solution is
\begin{equation}
\phi ({\mathbf k} \cdot {\mathbf r}) = - i {{C_1 \varepsilon_v} 
\over {\varepsilon_o}} e^{i \varepsilon_o ({\mathbf k} \cdot 
{\mathbf r}) / \varepsilon_v} + C_2\ .
\label{eq38}
\end{equation}

This shows that the fundamental equations for OM and DM are 
different and that they depend on different constants of Nature.  
As a result, the Schr\"odinger equation describes the quantum 
structure of OM, whereas the quantum structure of DM is 
described by the new asymmetric equation (see Eqs. \ref{eq37}
and \ref{eq38}).

\subsection{Implications for dark matter}

In the proposed fundamental equation for DM (see Eq. \ref{eq6} 
or Eq. \ref{eq21a}), the coefficient $C_w$ depends on the new 
universal constant of Nature $\varepsilon_o$, which represents 
the quanta of energy for DM.  The role of $\varepsilon_o$ is 
similar to the Planck constant but its physical meaning is different, 
namely, while the energy for OM is given by $E = \hbar \omega$, 
the energy for DM is simply $E_{o} = \varepsilon_o$.  In other 
words, the main difference between OM and DM is that 
{\it quantization} of DM can only be done using the quanta 
$\varepsilon_o$; this may explain the lack of observed radiation 
[4-9] and a significantly different behavior of DM than OM [23].

The obtained results imply that DM may be considered as a 
collection of DM elementary particles with mass $m$ and that 
these particles may exchange their energy by emitting or absorbing 
the quanta of energy $\varepsilon_o$.  Assuming that the particles
interact only gravitationally, the process of quanta exchange may 
produce a detectable gravitational radiation.

\section{Conclusions}

Among the two infinite sets of derived Galilean invariant equations,
the fundamental equations of matter are identified.  It is shown that 
in one set, the fudamental equation is the Schr\"odinger equation for 
ordinary matter, and in the other set, a new asymmetric equation is 
proposed to be the fundamental equation for dark matter.  The 
obtained results demonstrate that the equation for dark matter 
requires a new constant of Nature, which is called a {\it quanta of 
energy}.  The constant is only valid for dark matter and its physical 
implication is that quantization of dark matter energy is independent 
from frequency.  The resulting frequency independence may be 
responsible for the lack of electromagnetic radiation from dark 
matter, for its different behavior compare with ordinary matter, 
and it may also be responsible for the generation of detectable 
gravitational radiation.

{\bf Acknowledgment:} The author thanks Dora Musielak for valuable 
comments on the earlier version of this manuscript.  This work was 
partially supported by Alexander von Humboldt Foundation.


\end{document}